# Business and ethical concerns in domestic Conversational Generative AI-empowered multi-robot systems

Rebekah Rousi[1][0000-0001-5771-3528], Hooman Samani[2][0000-0003-1494-2798], Niko Mäkitalo[3][0000-0002-7994-3700], Ville Vakkuri[1][0000-0002-1550-1110], Simo Linkola[4][0000-0002-0191-589X], Kai-Kristian Kemell[4][0000-0002-0225-4560], Paulius Daubaris[4][0009-0008-4243-4751], Ilenia Fronza[5][0000-0003-0224-2452], Tommi Mikkonen[3][0000-0002-8540-9918] and Pekka Abrahamsson[6][0000-0002-4360-2226]

[1] University of Vaasa, Wolffintie 32, 65200 Vaasa, Finland
[2] University of Arts London, 272 High Holborn, London WC1V 7EY, United Kingdom
[3] University of Jyväskylä, Mattilanniemi 2, 40100 Jyväskylä, Finland
[4] University of Helsinki, Yliopistonkatu 4, 00100 Helsinki, Finland
[5] Free University of Bozen-Bolzano, Sparkassenstraße 21 - via Cassa di Risparmio, 21 39100, Bozen-Bolzano, Italy
[6] University of Tampere, Pohjoisranta 11A, 28100 Pori, Finland
`rebekah.rousi@uwasa.fi`

**Abstract.** Business and technology are intricately connected through logic and design. They are equally as sensitive to societal changes, and may be devastated by scandal. Cooperative multi-robot systems (MRSs) are increasing, entailing that robots of different types and brands can work together in diverse contexts. Conversational generative artificial intelligence (CGI) has dominated recent artificial intelligence (AI) discussions, with its capacity to mimic humans through use of natural language and production of media including deep fakes. Like MRSs, CGIs present huge potential in terms of revolutionizing processes across sectors - transforming the way humans do business. From a business perspective, cooperative MRSs alone, with potential conflicts of interest, privacy practices, and safety matters, beg for ethical examination. MRSs that are empowered by CGIs yearn for multi-dimensional and sophisticated methods to uncover imminent ethical pitfalls. The current study examines ethics in CGI-empowered MRSs, while reporting the stages of developing the MORUL model.

## 1 Introduction

Conversational Generative Artificial Intelligence (CGI) is riding the wave of fame (or infamy) at present times. No area of society is untouched by the breakthroughs made in the development of data-driven systems that are capable of communicating with humans via natural language. Areas particularly affected by this technology include business and robotics. Harnessing CGI in organizational operations can mean exponential business value [1]. Utilizing CGI in robotics means increased usability, accessibility and market potential of robotics [2]. Yet, embracing trending technological developments is not without risk. As major media headlines over the past few years have attested, the increasing occurrence of mishap in sophisticated data-driven systems, can have devastating consequences for humans, technology, and business alike. One of the main use contexts for these complex emerging products and services is the home. For instance, the global smart home market has been estimated to grow from 93.98 billion United States Dollars (USD) (2023) to 338.28 billion USD by the year 2030 [3]. In other words, market demand draws a landscape of complexity and multi-layered System of Systems (SoSs) into the private and often considered *sacred* space of the home [4][5]. Products such as refrigerators, vacuum cleaners, and toasters, become robots and (secret) public broadcasters [6]. Thus, ethics touches upon all levels of technological implementation in the home, because relationships between humans and objects have transformed [7].

CGI embedded Multi-Robot Systems (MRS) in domestic settings raise a number of ethical concerns for businesses [8][9]. The development of CGI-embedded MRSs has been largely industrial and business-oriented [10]. These systems have been designed to automate tasks and improve efficiency in a variety of industries, including manufacturing, healthcare, and customer service. As a result, the ethical considerations of CGI-embedded MRSs have often been overlooked. Businesses that develop or deploy CGI-embedded MRSs must carefully consider these ethical concerns and take steps to mitigate them. The present paper employs an applied ethics approach to investigate potential ethical concerns deriving from the development and deployment of data-driven multi-robot cooperative systems. By applied ethics, the authors refer to a case specific approach designed to how social ethical dilemmas practically unravel when particular technical, and social-technical (mixture of human and technological factors) are operationalized in certain contexts [11].

This present study concentrates on identifying potential ethical challenges in the development, deployment and implementation of multi-robot cooperative systems in the home. A scenario-based approach was applied to investigate the potential ethical concerns and moral implications of introducing heterogeneous multi-robots into domestic spaces. The research focused on the business value of isolating potential ethical problems before embarking on technical system development. Moreover, the study additionally aimed to operationalize the knowledge that not all ethical issues and related analyses can be undertaken pre-development. Rather, the ethical concerns and predicted stages at which analyses should be undertaken are divided according to the dimension of ethical concern (i.e., safety, security, societal concern) and components of these cybernetic systems.

## 2 Background

### 2.1 Large Language Models (LLMs) in Multi-Robot Cooperation

Large LLMs and CGI, are among some of the latest ML developments to gain widespread public popularity. In particular, OpenAI's Conversational Generative Pre-training Transformer architecture (ChatGPT) has been in headlines and public debate since approximately 2018 [1]. LLMs and CGI advance more recent trends in the popularity of chatbot development [12]. In the instance of chatbots, Natural Language Processing (NLP) is employed to communicate with users via searching for optimal responses offered by the information system. ChatGPT can be seen as an advanced form of chatbot that augments earlier versions of chatbots through the combination of deep learning and LLMs [13]. LLMs aim at predicting word sequences that are often utilized in human communication. Yet, this in itself engages processes that encourage bias or discrimination in algorithms as the processes rely on transformer architectures of neural networks as well as deep learning that depend on representative data [14]. ChatGPT, for instance, combines supervised fine-tuning with unsupervised pre-training to create answers that seem to be generated by human beings, and experts. This heightens both the social dimension of human-data interaction, as well as data accessibility to non-experts.

Currently, every prompt-based discussion with AI-based chatbots may be relatively costly, considering how many prompts are typically required for completing a single task, and how many people are using these models. Tech companies, such as OpenAI, Microsoft, Alphabet, and Meta, are aiming to take advantage of this emerging technology and trying to build business around AI-based technologies for personal and professional use. Due to the costs caused for training and running the models, companies are competing with different business strategies. For example, OpenAI provides its GPT model as a service via an API in order for new AI-based applications to be built on top of their models. On the other hand, new open source LLMs are being released on the Internet with varying capabilities and licenses. Meta presently provides its advanced LLAMA 2 model open source, and for limited commercial use.

Multi-robot cooperation entails that two or more robots, either of the same or other brands, models, or types work together to achieve common goals [15]. The goal of each robot may not be identical, however, there should be a common or higher goal among the robots that sees them cooperate towards a common outcome, i.e., that a home is safe and clean, or that services in a hospital are delivered in a timely and effective manner. The highest possible goal in this instance is the well-being of the human owner. One tenet of multi-robot cooperation is that it focuses on addressing complex operations that are nearly impossible to successfully carry out without a team [15][16]. At some level however, humans always exist in relation to the multi-robot processes. Whether programming or giving commands, or indeed, co-working with the robots, multi-robot cooperation should always be considered in relation to humans and the level of involvement they represent within various processes [17]. Consideration for human factors in co-working with multi-robot systems involves certain levels of complexity. Simões and colleagues [18] identified three levels of complexity: 1) the human operator and the technology itself; 2) recommendations and guidelines pertaining to the performance of human-robot teams affecting HRI; and 3) complex holistic approaches led by recommendations and guidelines that in turn affect HRI.

It may be assumed in any case that the point of departure in light of the human dimension in multi-robot cooperation is always a complex negotiation of completing systems, diverse operational goals, diverse corporate goals and strategies (played out within the logic of the systems), which is then influenced by standards, laws and recommendations. This means that the entry level to examining these types of systems always remains at Level 3 [18]. Given this, the process of preempting ethical issues already at the pre-development phase raises the investigation to the next level - Level 4 (systemic ethical forecasting in cybernetic systems). This forecasting entails understanding how the MRSs operate across human dimensions, one of which is communication [19]. Communication accounts for not only the functional capacity of humans to operate the multi-robot systems, but also the social-emotional components of HRI [19][21]. For this reason, CGI in forms such as ROSGPT or ChatGPT have had profound effects on the ways in which people interact with ML systems [20].

ROSGPT [22], introduces a pioneering approach that capitalizes on the immense potential of LLMs to propel human-robot interaction to unprecedented heights. This framework integrates ChatGPT into ROS2-based robotic systems, unveiling a synergy between language understanding and robotic control. Central to ROSGPT's advantages is its adept prompt engineering, leveraging ChatGPT's diverse abilities – from information elicitation to coherent train of thought – to convert unstructured natural language commands into precise, contextually relevant robotic instructions. ROSGPT harnesses LLMs' inherent learning capabilities, effortlessly eliciting structured commands from unrefined language inputs. The proof-of-concept demonstration, showcasing the translation of human language into actionable robotic directives, underscores ROSGPT's potential across varied applications. Beyond its immediate utility, ROSGPT's open-source implementation on ROS 2's platform not only fosters collaboration between robotics and natural language processing domains but also signals an exciting stride toward the realm of Artificial General Intelligence (AGI).

## 2.2 Business effects of AI Ethics, CGI, multi-robot cooperation

Ethics in the domains of AI have been hot topics for decades now, and this is becoming increasingly more so as AI is deployed widely in society. Earlier discussions applied the terms 'information ethics', 'machine ethics' and 'computer ethics' [23] to describe the field of examining ethical and moral implications of IT. Through engaging with applied ethics, scholars and practitioners are able to consider deeper philosophical questions through practical examples of how they manifest [24]. For instance, racism not only manifests through abuse and degradation, but also false accusation (see e.g., [25]). There is a sense of urgency spurred from the already emergent incidents involving machine learning (ML) technology utilization [23]. Whether the incidents involve matters of accountability and responsibility as witnessed in accidents in which human life has been harmed or damaged. The AI Incident Database [26] reported 90 incidents in 2022 alone, of AI-caused accidents, 45 already at the beginning of 2023. The rate of AI incidents seems to be increasing at a comparative pace to Moore's Law - doubling every year, similarly to the compounding capacity of computing speed [27]. These not only incur substantial costs in damages and potential insurance premiums, but pose serious problems from basic issues of human respect, safety, and dignity, to the severe tarnishing of reputation for businesses who do not embrace humane factors as a part of their data-driven business strategy [28].

The 2018 self-driving Uber accident in which a pedestrian was fatally wounded (see e.g., [29]) incurred irreparable immaterial damage. This no doubt contributed to loss of income, hindered self-driving vehicle development (and brands), tarnished Uber (now owned by Aurora Innovations) as a transportation service, and the operator who was responsible for monitoring the vehicle. While the human operator has been found guilty of negligence, the repercussions of the accident in terms of legal expenses and loss of consumer trust are remarkable. Not only were the direct implicated actors affected, but the US Federal Government was also accused of not properly regulating the industry. Moreover, had the accident led to a total abandonment of self-driving vehicles by companies such as Uber, profit trajectories would be thrown off course, because drivers account for 80% of all costs - self-driving units being evaluated at 7 billion United States dollars already in 2020 [29].

Business intercedes on many dimensions of AI and robot ethics. From privacy-related issues and dark practices of the surveillance economy, to platform economy logic, and 'login - lockin' cultures, business needs to be considered from both back and front-end perspectives. When it comes to ethics, business itself can be its own worst enemy. The logic that may pave the way to patents and trade secrets, may be guilty of fostering ethical potholes such as black box systems diminishing customer and user trust, and even simply, bad user experience with greater social repercussions. The dance between ethics and business is like a temptation-filled devil's tango. The appeal of fast profits blinds many of careful foresight in business strategy. Effective management of ethics in AI and robotic development would not just mean better business strategy, but also longevity [30].

## 3    Method

In the present study the researchers employed a qualitative explorative method via two workshops. A scenario-based approach was used to contextualize the inquiry that entailed imagining that several robots of different use purpose, brand and type, utilizing CGI technology were implemented in the home (see Fig. 1). In the scenario, two cleaning robots of the same brand and make have been used in the home for quite some time. The new addition of a robot arm from a different brand and manufacturer elicits ethical concerns when considering the need for all robots to cooperate in order to perform tasks to reach certain goals. The workshops were held at separate times: Workshop 1 (W1) was held during February, 2023, for two days face-to-face at a lab hosted by one of the participating research institutions; and Workshop 2 (W2) was held in June, 2023, for one hour via Zoom. The idea behind the separate timing was to allow for the analysis of W1 results, in order to synthesize and construct a preliminary framework for W2. The preliminary framework was seen as the basis for modeling a matrix that eventually will serve as a scaffolding for ethical multi-robot development. The matrix would include facets starting from ethical business strategy (understanding the influence of economic superstructures in molding the logic of technological products), to hardware and software, human-technology interaction, larger societal repercussions, and back again to business impact.

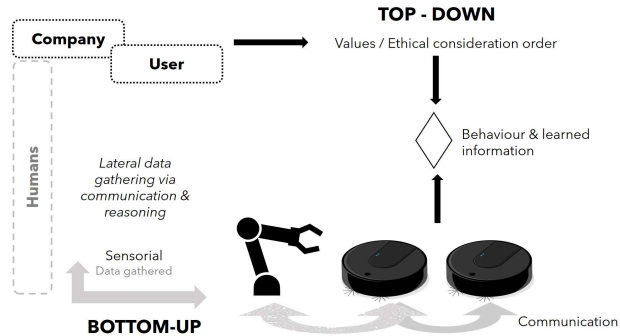

**Fig. 1.** Domestic scenario of two cleaning robots and one robot arm - understanding relations between layers and domains of multi-robot cooperation from a techno-corporate perspective

Qualitative data was collected in the form of brainstorming drawings and notes. The material from W1 was originally in paper versions, which were subsequently photographed and digitally archived. The material from W2 was produced on Google Jamboard. During processing of the data - transferal from the drawing boards to excel and image files - preliminary thematic categories were established. Extra rounds of thematic analysis [31] were performed by the research team in an excel document. The study was conducted via a constructivist grounded theory [32] approach in order to build on previous AI ethics principles, guidelines and methods (see e.g., [23], while allowing for deeper examination of specific details and dimensions that are phenomenologically unique to the domain of multi-robot cooperation.

### 3.1   Ethical and responsible research

As this is a novel space of research that deals with ethics across a range of levels, from basic practical levels to higher levels of abstraction, the research team deemed the safest and most responsible approach to be that of internal inquiry. To avoid physical or psychological harm, the team of experts maintained the empirical component outside the realm of physical human-robot or robot-robot interactions. Rather, the researchers deliberated through discussion, illustration and writing. All researchers involved in the workshops were willing participants, agreeing the use of their data, exercising scholarly agency as experts within their respective fields. In compliance with the General Data Protection Regulation (GDPR), all data is stored in secure password-protected digital locations to which only two main researchers have access. No personal data is stored with the research data.

### 3.2   Participants

Each workshop comprised eight participants, rendering $N$=16 contributions in total. Five participants participated in both workshops ($N$=10 contributions) while six par-ticipants only participated in one of the workshops. This meant that the overall total of individual participants was $N$=11. All participants possessed a higher tertiary degree, starting at PhD level researchers and higher. The gender distribution was two females and nine males. The fields of expertise that the participants represent are: software engineering and computer science; robotics and software for robotics; edge intelligence; computing education; information systems; cognitive science; human computer interaction; communication; and social ethics.

### 3.3 Procedure

The workshops were planned and agreed upon in a series of online meetings. In these meetings the strategy was deliberated, goals were set, as well as timing, procedure and locations were established. The context for the scenario was decided upon via several brainstorming sessions in which the team examined areas, environments and situations in which ethics and moral conduct would be considered as most sensitive [5]. After identifying several domains including education, healthcare, elderly care, and the home, the team selected the home, both for its intimate framing of privacy, as well as its diversity [4]. While there are central features defining a home - living space, kitchen, bedroom etc. - the ways in which people appropriate, populate, and utilize their spaces is quite eclectic [4]. This is as opposed to public institutions such as hospitals that are laden with rules, standards and top-down regulations.

**Workshop 1**

Workshop 1 took place in person, on location at the lab of one of the participating research institutions. The lab is designed as an innovation space with a central meeting area equipped with audio-visual and teleconferencing equipment, as well as traditional tools such as flipcharts, post-it notes, colored pens. One participant contributed via Zoom for logistical reasons. The workshop was held over a two-day interval. The procedure entailed a round of introductions and articulating our interests in relation to the topic for the participants who had not been involved in the previous online planning sessions. The workshop proceeded as seen in Table 1.

**Table 1.** Workshop 1 procedure.

| Step | Stage | Description |
|---|---|---|
| 1 | *Re-cap of use context and scenario* | Narrative unfolds in the home. Two similar robots (vacuum cleaners) and a newly introduced robot arm |
| 2 | *Independent mind-mapping of ethical concerns [unstructured]* | Independent work (30 min.), focus on ethical concerns |
| 3 | *Group discussion and comparison of findings* | Discussion of mind-maps, sharing ideas and introducing new concerns that arose in the group discussion |
| 4 | *Identification of the layers* | Identifying layers implicated in LLM-enabled multi-robots |
| 5 | *Model formulation* | Deliberation of actionable models of ethics in multi-robot collaboration that could be utilized within the |

programming process

**Workshop 2**

Workshop 2 was carried out via Zoom to allow for international collaboration while some members of the study were traveling. The duration of the workshop was two hours and held on Google Jamboard. Building on the findings of Workshop 1, Workshop 2 was structured according to a matrix of multi-robot cooperation domains and layers: Human-Interaction; Sensorial Layer (robot hardware); Deliberation (robot brain); Behavioral (robot hardware); Communication and Networking (robot-to-robot interaction); and System of Systems (network or systems). From the human perspective, considerations of ethical aspects were encouraged to be thought of through the frames of: 1) safety, 2) security, and 3) societal dimensions. The procedure of Workshop 2 is observed in Table 2.

**Table 2.** Workshop 2 procedure.

| Step | Stage | Description |
| --- | --- | --- |
| 1 | *Instructions & breakdown of procedure + use-context re-cap* | Use context is the home and workshop members are encouraged to think of all potential ethical issues and scenarios arising from the introduction of LLM-powered multi-robot cooperation in domestic spaces |
| 2 | *Independent mind-mapping of ethical concerns [unstructured]* | Independent work (30 min.), focus on ethical concerns |
| 3 | *Group discussion & comparison of findings* | Groups progressed through the domains and layers of multi-robot cooperation as well as the human dimensions of the concerns |
| 4 | *Layer and domain refinement* | Group reflected on the earlier version of the layers and domains based on new findings arising in W2 |
| 5 | *Model refinement* | MORUL model for ethical CGI-enabled multi-robot development further refined |

### 3.4   Analysis

Thematic analysis [31] was employed to analyze the data of both workshops. In the case of Workshop 1, the researchers transcribed mind-maps, notes and illustrations that had been expressed on large flip chart sheets into excel sheets. From Workshop 2, the the Google Jamboard notes were transferred into excel. The analysis took place in three steps: 1) sorting data into themes; 2) refining the themes; and 3) performing frequency analysis to determine which themes arose in relation to which layer of the multi-robot systems. The themes were compared between both data sets, and cross-validated among the research team to ensure consensus of the themes and labels. The themes were again reviewed according to the technological layers, as well as the domains (i.e., safety, security, and society) that they are implicated with. The business dimension of the multi-robot ethical concerns has been positioned as a superstructure

(economic and logic base) during and after analysis to make sense of the influence that corporate competition through technological design has on the ethical implications from conceptualization to implementation of the multi-robot systems.

## 4 Results

In total, 21 themes arose from the data. The themes and their quantities varied from Workshop 1 (W1) to Workshop 2 (W2). In W1, the emergent themes from 61 constructs (expressions) were: data security and privacy (3 - 4.9%); corporate dominance (3 - 4.9%); communication (17 - 27.9%); cooperation (10 - 16.4%); reliability and recover (1 - 1.6%); logic and standards (2 - 3.3%); human oversight (5 - 8.2%); prioritization/hierarchy (2 - 3.3%); trustworthiness/virtue (5 - 8.2%); executive function (2 - 3.3%); maleficence (3 - 4.9%), user experience (UX, 6 - 9.8%); and legislation (2 - 3.3%). The distribution of frequencies can be seen in Figure 2.

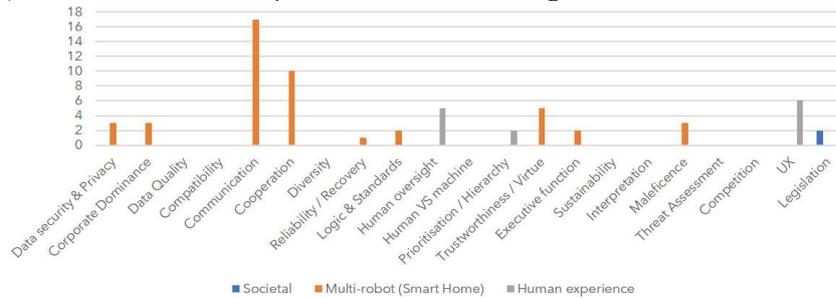

**Fig. 2.** Frequencies of ethical concerns expressed in Workshop 1

All themes in addition to the legislation theme are displayed in Figure 1. Based on the percentage of frequencies, *communication* (27.9%) was by far the most mentioned theme. Attributes associated with communication included communication failure between brands and makes of robot - corporate strategy and/or mere incompatibility. Communication was additionally connected to maleficence in cases whereby robots of competing companies may deliberately offer each other misleading communication. Another concern raised in relation to communication was the potentiality for a black box scenario in which human users, via CGI, communicate on one level with the robots, yet the robots themselves communicate and operate on a different level to humans. This may lead to various aspects of data collection and sharing of data that human users are unaware of. Following communication is *cooperation* (16.4%). Both through communication as well as strategic behavior, robots may either withhold crucial information and task sharing from one another, placing obstacles in robots of competing brands' pathways (including themselves). While these tactics may seem childish, one may only look towards current and recent world leaders to understand that people (and companies) will do anything to ensure an advantage over competition. Thus, other thematic aspects can be seen as related to (*corporate dominance*, *trustworthiness / virtue*, and *maleficence*), intertwined with (*prioritization / hierarchy*, *executive function*, *legislation*, *logic & standards*), and resulting from (*UX*, *human* *oversight* and *data security & privacy*) ethical concerns in *communication* and *cooperation*.

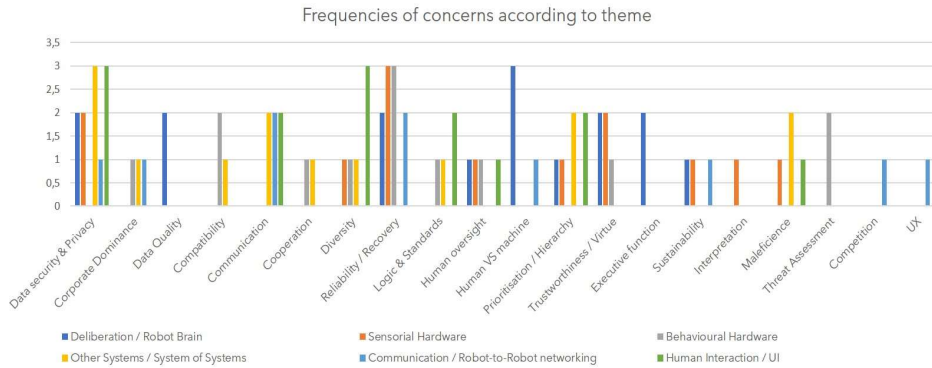

**Fig. 3.** Frequencies of ethical concerns expressed in Workshop 2

W2's results follow a factor logic that connects the themes strongly to related domains or layers (see Figure 3). Thus, issues of *diversity* (8 - 10%) including matters of accessibility and linguistic input preference (capabilities) were mentioned mostly in relation to the layer of human interaction. Diversity was also mentioned in reference to the sensorial hardware, other systems and behavioral hardware, and these can be understood as intertwined with the *communication* theme. While *communication* was mentioned six (7.5%) times in reference to other systems, robot-to-robot networking, and human interaction, other themes rose to the fore. *Interpretation* (1 - 1.3%) resonates with communication, and was mentioned in conjunction with the sensorial hardware. *Human versus machine* (4 - 5%) manifested in comments regarding the logic of deliberation / robot brain and communication / robot-robot networking. Perhaps related to the theme of *human oversight* (4 - 5%) and the ability of humans to keep pace of what is happening within the systems, and as such, maintain a certain level of control *human versus machine* radiates an element of techno-paranoia and the prospect of developing systems that eventually humans may not be able to control. *Logic & standards* (4 - 5%) were mentioned in relation to the system of systems, behavioral hardware layer, as well as the human interaction layer. These may be seen as both enablers of CGIs in multi-robot cooperation (standardizing and coordinating cooperation between and across robots, with humans), and gray areas when considering built-in logic that differs across language boundaries, and standards.

The *executive function* (2 - 2.5%), was noted and linked to the robot brain, which should not be surprising. Yet, in relation to this layer, there were thoughts that could be connected to the *human versus machine* theme, as well as *trustworthiness & virtue* (5 - 6.3%). This is considered from the perspective that the goals, and hierarchy of goals guided by the executive function could very easily be dictated by corporate objectives rather than the concerns of human users. *Maleficence* was mentioned more (4 - 5%) in relation to other systems, yet was also connected to the sensorial hardware and human interaction domains. This theme connected with the intention of the company or developer (for instance, the Amazon ownership of Roomba was raised often in discussion) and reasons for particular types of ownership in light of potential data collection, data sharing (sales), and 'lock-ins' (need to be locked/logged into certain systems at all times). *Sustainability* (3 - 3.8%) was a theme connected to the deliberation / robot brain layer, sensorial hardware, and robot-to-robot networking. Issues of programmed obsolescence and consideration for corporate responsibility in relation to the production of components, as well as recycling and disposal of non-working devices were raised.

The results led to the deliberation of a diagram that organized themes in relation to how they were represented within the workshops (see Fig. 4). The authors of the current paper acknowledge the role of culture in shaping not only society, but all the socio-technical and corporate aspects of any technological development. This said, the *cultural* domain is nestled next to the *systems and artefacts* domain due to their interwoven relationship that spans from tribal rituals and hand tools to complex AI and multi-robot systems. The *societal domain* is seen here as a holistic framework that is characterized by standards, regulations and general governance. As mentioned earlier, the researcher workshop participants were highly critical regarding the effectiveness of current regulatory frameworks (including the recently released draft of the EU AI Act, see [33] as it seems that the development is by far outpacing the speed of governance [ref.] over the technology in society.

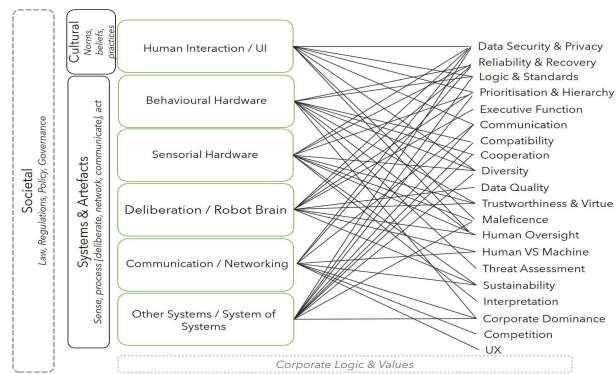

**Fig. 4.** Organization of domains, layers and themes

The layers are subsequently arranged from the 'top' layer of human interaction or user interface (UI) layer to the behavioral hardware - the observable action layer that both undertakes tasks and interacts with humans. Both processes and layers are interwoven and interdependent - they are SoSs. CGI was interpreted as the buffer between non-expert humans and functionality. It is not simply a UI component in itself, yet provides a substantial logic that feeds into the SoSs via provision of training data collected from users, cross-robot communication (additionally with robots or bots not directly present within the domestic setting), and above other things, has the capacity to establish affinity between human beings and robots through its seeming intelligence.

The behavioral hardware is more directly attached to the understanding of the robot unit's actions. However, as understood in the case of adding CGI, more than one unit is already present within the seemingly single-standing robot. Sensorial hardware, while embedded within the physicality of the robots, also connects with what we can understand as the 'robot brain' - central processing unit utilized for deliberation. Once again, this lends to gray area territory due to the interconnected nature of the robots with similar, and also *other* robots. The SoS entails the complex systems supporting the robots, yet additionally connects with the broader system of domains (societal, artifactual, and corporate). Figure 4 sheds light on the thematic findings of the workshops in respect of the layers they predominantly attach with.

## 5    Discussion

The use of CGI-embedded MRSs in domestic settings raises a number of ethical concerns for businesses. The development of CGI-embedded MRSs has been largely industrial and business-oriented, with little pre-, and during production thought for the ethical implications of MRSs and design choices [10][22]. These systems have been designed to automate tasks and improve efficiency in a variety of industries, including manufacturing, healthcare, and customer service. As a result, the ethical considerations of CGI-embedded MRSs have been largely overlooked. Businesses that develop or deploy CGI-embedded MRSs must carefully consider a range of ethical concerns, and their impact across domains, including safety, security, liability and accountability, society, and naturally, the business itself.

While the field of human-computer interaction chants about considering everything and everyone already at the beginning, this research shows that not all ethical issues can be accounted for at the conceptualization phase. For example, the ethical concerns of social media platforms were not fully understood until after they were widely adopted. CGI-embedded MRSs are no different. It is likely that ethical concerns regarding these systems will not be fully understood until they are widely deployed. It is possible that CGI-embedded MRSs could be used to spread misinformation or propaganda, or to discriminate against certain groups of people. In knowing the *unknown*, careful business strategy involves preempting the stages (chronological), and levels (components, domains and potential impact) in which ethical issues may arise, or at best, should be assessed. For instance, if a matter of concern may be bias due to LLM training data, several LLMs should be adopted within the systems. If ML in the back-end of the robots can be expected to occur rapidly, check-points, communication protocols, and 'pit-stops' (points at which the systems stop) should be built in to ensure that humans – general users and experts – can observe and understand what is happening within the learned data, instilling transparency and human oversight. There are a number of other actionable points and operations that businesses and developers alike can predict for intervention and management, such as data offloading.

### 5.1    Limitations

The current study presents a number of limitations. Firstly, the empirical study presents a conceptual scenario-based investigation of CGI-empowered MRSs in the home. There was a limited number of participants, and the expert sample could have been strengthened with more research from the disciplines of ethics, law, software engineering and robotics, as well as psychology. Future steps would entail including experts from these disciplines, in addition to delving more specifically into the traits and problematics that CGI pose for MRSs – deep fakes and anthropomorphism are two areas that challenge the ethical use of CGI by its very nature. May people see Britney Spears or their *favorite* neighbor sweeping their floors any time soon? Where are the boundaries and/gray areas of privacy and intellectual property concerns when personalizing personal consumer CGI-empowered MRSs? Other limitations include the fact that this study to date has almost strictly focused on front-end issues, ignoring the back-end realm in which matters such as accuracy can severely impinge on the operations of the systems. In turn, the corporate influence and affects multiple LLMs defining the logic of the systems need to be critically examined.

# 6   Conclusion

As for long-term strategy, social responsibility and corporate reputation, businesses should develop clear policies and procedures that preempt and avoid foreseeable issues already at the strategy phase of innovation. This includes instilling transparency and clarity regarding of privacy policies and practices, as CGI-empowered MRSs are constantly collecting, utilizing and disclosing data. By addressing these ethical concerns, businesses further ensure that CGI-embedded MRSs are used in responsible and ethical ways, potentially preventing incidents that cost business and society millions if not billions in damages. Indeed, ethical coverage of CGI-empowered MRSs may be worth billions in added-value.

It is important to start considering the ethical implications of CGI-embedded MRSs now, before they are widely deployed. This will help ensure that these systems are used in a responsible and ethical manner. Steps must be taken to mitigate ethical issues. Yet, the timing and level upon which mitigation takes place varies according to the nature of the concern itself, its cause, and how it manifests within the systems. Ethics permeates the entire hardware and software development process from design to operations. It is far cheaper to make changes during design and far more expensive, and maybe even nigh impossible, to fix ethical issues in production. While issues like bias can be may be tackled with model re-training that can be done even after deployment, if the goal or purpose of the system itself is the problem (e.g., social credit scoring with facial recognition on the streets), it may be very hard to tackle – due to its short-term business value (i.e., attractiveness for places and business such as airports).

In terms of practical implications, the issues already identified within this paper may form the platform upon which organizations may be guided. In particular, the MORUL framework for ethical multi-robot cooperation has its basis in the dual process presented in the workshop scenario method reported here. The authors would also like to emphasize two fundamental challenges that AI ethics per se, repeated face: 1) a lack of consensus regarding what AI and AI-robot ethics *is* – requiring a

14framework to generate broad shared understanding among communities; and 2) *how* to engage in AI, and AI-robot ethics – how can attributes such as fairness, transparency, and privacy etc. be instilled in data-driven systems? Once more, a framework is needed. Future papers will document the progress of MORUL, and will present its application with working demos and prototypes. At this time, we may consider MORUL as a *call to action* to gear business up for considering ethical issues from the outset, as a part of best practice, and as an *essential* salespoint.

## References

1. Dwivedi, Y. K., Kshetri, N., Hughes, L., Slade, E. L., Jeyaraj, A., Kar, A. K., ..., Wright, R.: "So what if ChatGPT wrote it?" Multidisciplinary perspectives on opportunities, challenges and implications of generative conversational AI for research, practice and policy. International Journal of Information Management, **71**, 102642 (2023) DOI: 10.1016/j.ijinfomgt.2023.102642
2. Stige, Å., Zamani, E. D., Mikalef, P., & Zhu, Y. Artificial intelligence (AI) for user experience (UX) design: a systematic literature review and future research agenda. Information Technology & People. Vol. **ahead-of-print** No. (2023). DOI: 10.1108/ITP-07-2022-0519
3. Fortune Business Insights, The global smart home market size was valued at $80.21 billion in 2022 & is projected to grow from $93.98 billion in 2023 to $338.28 billion by 2030, https://www.fortunebusinessinsights.com/industry-reports/smart-home-market-101900, last accessed 2023/09/10
4. McClain, L. C.: Inviolability and privacy: The castle, the sanctuary, and the body. Yale JL & Human., **7**, 195 (1995)
5. Lutz, C., Newlands, G.: Privacy and smart speakers: A multi-dimensional approach. The Information Society, **37**(3), 147–162 (2021) DOI: 10.1080/01972243.2021.1897914
6. Yao, Y., Basdeo, J. R., Mcdonough, O. R., Wang, Y.: Privacy perceptions and designs of bystanders in smart homes. Proceedings of the ACM on Human-Computer Interaction, 3(CSCW), 1–24 (2019) DOI: 10.1145/3359161
7. Atlam, H. F., Wills, G. B.: IoT security, privacy, safety and ethics. Digital twin technologies and smart cities, 123–149 (2020) DOI: 10.1007/978-3-030-18732-3
8. Nascimento, N., Alencar, P., Cowan, D.: Self-Adaptive Large Language Model (LLM)-Based Multiagent Systems. arXiv preprint arXiv:2307.06187 (2023).
9. Authors et al.: I Trust You Dr. Researcher, but not the Company that Handles My Data – Trust in the Data Economy. In: Proceedings of the Hawaii Conference on System Science (HICSS) 2024 (Forthcoming)
10. Mandi, Z., Jain, S., Song, S.: RoCo: Dialectic Multi-Robot Collaboration with Large Language Models. arXiv preprint arXiv:2307.04738 (2023). DOI: 10.48550/arXiv.2307.04738
11. Beauchamp, T. L. The nature of applied ethics. A companion to applied ethics, 1–16. In R. G. Frey, Christopher Heath Wellman (Eds.) A companion to applied ethics (2003)
12. Lokman, A. S., Ameedeen, M. A.: Modern chatbot systems: A technical review. In: Proceedings of the future technologies conference, pp. 1012–1023 (2018)
13. Radford, A., Narasimhan, K., Salimans, T., Sutskever, I.: Improving language understanding by generative pre-training. Preprint. 1–12 (2018).
14. Vaswani, A., Shazeer, N., Parmar, N., Uszkoreit, J., Jones, L., Gomez, A. N., ... Polosukhin, I.: Attention is all you need. Advances in neural information processing systems, **30** (2017)